\documentclass[paper]{myJHEP3}
\usepackage{amsmath}
\usepackage[english]{babel}
\usepackage{cite}
\newcommand\mur{\mu_{\sss\rm R}}
\newcommand\sss{\scriptscriptstyle}
\newcommand{\bqa}{\begin{eqnarray}}
\newcommand{\eqa}{\end{eqnarray}}
\newcommand{\nl}{\nonumber \\}
\newcommand{\dbar}{\bar D}
\newcommand{\qbar}{\bar q}
\title{Integration-by-parts identities in FDR}
\author{Roberto Pittau\\
        Departamento de F\'isica Te\'orica y del Cosmos and CAFPE,
  Campus Fuentenueva s.n., Universidad de Granada, E-18071 Granada, Spain \\
  E-mail: \email{pittau@ugr.es}}
\abstract{
Four-dimensional renormalized (FDR) integrals play an increasingly important role in perturbative loop calculations. Thanks to them, 
loop computations can be performed directly in four dimensions and with no ultraviolet (UV) counterterms. In this paper I prove that integration-by-parts (IBP) identities can be used to find relations among multi-loop FDR integrals. Since algorithms based on IBP are widely applied beyond one loop, this result represents a decisive step forward towards the use of FDR in multi-loop calculations.}
\preprint{}
\begin{document}
\section{Introduction}
\label{intro}
When computing loop corrections in a quantum field theory (QFT) described by a Lagrangian ${\cal L}$, UV infinities appear that have to be  dealt with in a mathematically consistent way.
The customary approach~\cite{Collins:1984xc} accomplishes this in two steps. First, the  UV divergent loop integrals are regularized~\cite{'tHooft:1972fi}. Then, the dependence on the regulator is eliminated by re-absorbing it --\,order by order in the perturbative expansion\,-- in the parameters of ${\cal L}$~\cite{Bogoliubov:1957gp,Hepp:1966eg,Zimmermann:1969jj}. 

In~\cite{Pittau:2012zd} it has been shown that it is possible to {\em define} the loop integration avoiding, from the very beginning, the occurrence of UV divergences. This new type of integration, called FDR, \footnote{Acronym of Four Dimensional Regularization/Renormalization.} does not depend on any UV cutoff and coincides with the usual integration in the case of UV convergent integrals. Moreover, it preserves the algebraic manipulations of the integrands 
and their shift invariance properties needed to prove the Ward-Slavnov-Taylor identities of the QFT at hand. In this way, all symmetries of ${\cal L}$, including gauge invariance, are preserved and the QFT gets renormalized by simply interpreting the loop integrals as FDR ones.\footnote{A {\em finite} renormalization is only necessary to link the parameters of ${\cal L}$ to physical observables.} The main advantage of FDR is that loop calculations can be carried out directly in the physical four-dimensional space and without re-absorbing UV infinities in the Lagrangian. As a consequence, Feynman diagrams containing counterterms are absent. In~\cite{Pittau:2013qla,Donati:2013iya,Donati:2013voa} the FDR strategy has been successfully applied to compute several processes at the one- and two-loop
accuracy in renormalizable QFTs.\footnote{For a physical interpretation of FDR and its possible use in non-renormalizable QFTs see~\cite{Pittau:2013ica}.}

The complexity of the multi-loop calculations is such that algorithms
are needed to reduce the problem to a small set of loop integrals, called master integrals (MI). 
In dimensional regularization (DR) one of the most powerful techniques is the use of relations among different integrals deduced via IBP identities~\cite{Tkachov:1981wb,Chetyrkin:1981qh}. For a given set of integrals to be solved, one picks an integral and generates an IBP identity, which solves for the most difficult integral, and so on~\cite{Laporta:2001dd}.  Thus, in a typical case, $10^4-10^6$ integrals can be expressed in terms of ${\cal O}(10-100)$ MIs~\cite{Marquard:2008qn}.  As for the actual computation of the MIs, IBP relations also serve as a starting point to write down linear systems of first-order differential equations in the kinematic invariants, that can be used for the determination of their analytic expressions~\cite{Kotikov:1990kg,Remiddi:1997ny,Gehrmann:1999as,Argeri:2007up,Smirnov:2012gma,Papadopoulos:2014lla}. Furthermore, methods exist to cast these systems in a canonical form in which all the analytical properties are explicitly exposed and the solution can be determined algebraically~\cite{Henn:2013pwa,Argeri:2014qva,DiVita:2014pza}.

Due to the algebra-preserving properties of their definition, FDR loop integrals can be reduced to MIs via algebraic procedures at the integrand level. For instance, the Passarino-Veltman~\cite{Passarino:1978jh} or OPP~\cite{Ossola:2006us} algorithms, both based on tensor reduction, hold in FDR.
While tensor reduction is sufficient at one-loop --\,where the set of MIs is known\,-- more sophisticated methods are needed, as discussed, to identify and compute the MIs at two-loops and beyond. Thus, it is crucial to establish whether procedures based on IBP identities can be formulated in the context of FDR. 

In this paper, I prove that four-dimensional IBP identities {\em do hold} under the FDR integral sign. Therefore, algorithms that make use of integration-by-parts techniques to compute multi-loop amplitudes are allowed in FDR.
Moreover, owing to the four-dimensionality of FDR, simpler formulations are expected compared with IBP procedures relying on DR.

The paper is organized as follows. In Sec.~\ref{FDR} the definition of FDR integration is reviewed. Section~\ref{IBP} describes the origin of the IBP relations. Section~\ref{IBPinFDR} illustrates the use of IBP in the context of FDR and Appendix~\ref{appb} collects two explicit examples.
\section{The FDR integration}
\label{FDR}
In FDR the UV subtraction is encoded in the {\em definition} of the loop integration by means of a twofold procedure.
First the $+i 0$ propagator prescription is identified with a mass $\mu^2$, such that $+i0 = -\mu^2$.\footnote{An imaginary part 
compatible with the $+ i 0$ prescription is given to $\mu^2$ itself. Furthermore, unlike in DR, the limit $\mu \to 0$ is taken {\em outside} integration.}
Secondly, the UV divergent terms are subtracted at the integrand level.\footnote{FDR integration and normal integration coincide in UV convergent integrals because  no integrand has to be subtracted in this case.}
This definition produces finite and regulator independent loop integrals that maintain the mathematical features needed to preserve the original symmetries of the QFT.
In this Section I illustrate FDR and its main properties with the help of simple examples. 

Consider the one-loop integrand
\bqa
\label{eq:1loopFDR}
\frac{1}{(\qbar^2-M^2)^2}
\eqa
with
\bqa
\label{eq:split}
\qbar^2 \equiv q^2 -\mu^2.
\eqa 
The corresponding FDR integral
is {\em defined} as 
\bqa
\label{eq:1loopFDRint}
\int [d^4q] \frac{1}{(\qbar^2-M^2)^2} &\equiv& \lim_{\mu \to 0}
\int_{\rm R} d^4q \left(
\frac{1}{(\qbar^2-M^2)^2} -\left[\frac{1}{\qbar^4}\right]
         \right) \nl
&=& \lim_{\mu \to 0} \int d^4q \left(
   \frac{M^2}{\qbar^4(\qbar^2-M^2)}
 + \frac{M^2}{\qbar^2(\qbar^2-M^2)^2}
         \right),
\eqa
where the subtracted term is conventionally written between square brackets
and $\int_{\rm R}$ denotes the use of an arbitrary UV regulator ${\rm R}$. ${\rm R}$ is needed because the two integrands in the first line of Eq.~(\ref{eq:1loopFDRint})  
are separately UV divergent, although their difference is not.
The UV dependence can be explicitly canceled by using partial fractioning
\bqa
\label{eq:partial}
\frac{1}{\qbar^2-M^2} = \frac{1}{\qbar^2} + \frac{M^2}{\qbar^2(\qbar^2-M^2)},
\eqa
that gives the second line of Eq.~(\ref{eq:1loopFDRint}), in which ${\rm R}$ can be dropped\footnote{Thus, FDR integrals {\em do not depend on any specific UV regulator.}}.

The rule to construct the subtraction integrands --\,such as $\left[\frac{1}{\qbar^4}\right]$\,-- is that they are allowed to depend on $\mu^2$ {\em but not on physical scales}.
In practice, they are automatically generated by expanding the original 
integrands by means of Eq.~(\ref{eq:partial}).  E.g.
\bqa
 \label{eq:1loopex}
\frac{1}{(\qbar^2-M^2)^2} = 
\left[\frac{1}{\qbar^4}\right]
 + \frac{M^2}{\qbar^4(\qbar^2-M^2)}
 + \frac{M^2}{\qbar^2(\qbar^2-M^2)^2}
\eqa
directly produces Eq.~(\ref{eq:1loopFDRint}).

Notice that $\mu^2$ serves as a temporary infrared (IR) regulator. Indeed, although the original integral is free of IR singularities, subtracting $\left[\frac{1}{\qbar^4}\right]$ creates a
$\ln \mu^2$ that forbids one to directly set $\mu$ to zero in Eq.~(\ref{eq:1loopFDRint}).
However, $\mu$  can be traded for an arbitrary scale $\mur$\footnote{Interpreted as the renormalization scale.} by observing that  the IR logarithm originates from the $q^2 \sim 0$ integration region of 
\bqa
\label{eq:divint}
\lim_{\mu \to 0} \int_{\rm R} d^4q \left[\frac{1}{\qbar^4}\right].
\eqa
Therefore, its coefficient is {\em independent} of ${\rm R}$,\footnote{As an example,  when evaluated in
$n= 4 + \epsilon$ dimensions,  Eq.~(\ref{eq:divint}) reads
\bqa
\label{eq:DR}
 \lim_{\mu \to 0}  \mur^{-\epsilon} \int d^nq \left[\frac{1}{\qbar^4}\right]
= 
i \pi^2 \lim_{\mu \to 0} \left(\Delta-\ln \frac{\mu^2}{\mur^2}\right)
+ {\cal O}(\epsilon), 
\eqa
where
\bqa
\Delta= -\frac{2}{\epsilon}-\gamma_E -\ln \pi,
\eqa
while the use of a hard cutoff $\Lambda_{\rm UV}$ gives
\bqa
\label{eq:hard}
\lim_{\mu \to 0} \int_{\Lambda_{\rm UV}} d^4q \left[\frac{1}{\qbar^4}\right]
=
-i \pi^2 \lim_{\mu \to 0} \left(1+\ln \frac{\mu^2}{\mur^2}
                               +\ln \frac{\mur^2}{\Lambda_{\rm UV}^2}\right),
\eqa
with $\Lambda_{\rm UV} \to \infty$.
} so that Eq.~(\ref{eq:1loopFDRint}) can be {\em re-defined}, in a regulator independent way, by sidestepping the subtraction of $\ln \frac{\mu^2}{\mur^2}$, where $\mur$ separates $\ln \mu^2$ from the rest. As a consequence, $\lim_{\mu \to 0}$ can now be taken and Eq.~(\ref{eq:1loopFDRint}) develops a dependence on $\mur$.\footnote{Explicitly, if $\ln \frac{\mu^2}{\mur^2}$ in Eq.~(\ref{eq:DR}) 
{\em is not subtracted}, one obtains
\bqa
\int [d^4q] \frac{1}{(\qbar^2-M^2)^2} = 
i \pi^2 \lim_{\mu \to 0} \left(\Delta-\ln \frac{M^2+\mu^2}{\mur^2}-\Delta
\right)= -i \pi^2  \ln \frac{M^2}{\mur^2},
\eqa
that coincides with the results one derives from Eq.~(\ref{eq:hard})  
\bqa
\int [d^4q] \frac{1}{(\qbar^2-M^2)^2} = 
-i \pi^2 \lim_{\mu \to 0} 
\left(
 1+\ln \frac{M^2+\mu^2}{\Lambda_{\rm UV}^2}
-1-\ln \frac{\mur^2}{\Lambda_{\rm UV}^2}
\right)= -i \pi^2  \ln \frac{M^2}{\mur^2}.
\eqa
}

Tensors are defined likewise. For instance
\bqa
\label{eq:tensor}
\int [d^4q] \frac{q^\alpha q^\beta}{(\qbar^2-M^2)^3} &\equiv& \lim_{\mu \to 0}
\int_{\rm R} d^4q \left(
\frac{q^\alpha q^\beta}{(\qbar^2-M^2)^3} -\left[\frac{q^\alpha q^\beta}{\qbar^6}\right]
         \right) \nl
&=& \lim_{\mu \to 0} \int d^4q\, q^\alpha q^\beta \left(
   \frac{M^2}{\qbar^6(\qbar^2-M^2)}
 + \frac{M^2}{\qbar^4(\qbar^2-M^2)^2}
 + \frac{M^2}{\qbar^2(\qbar^2-M^2)^3}
         \right).\nl
\eqa

Special care is necessary when Feynman rules generate integration momenta squared $q^2_i$ in the numerator. Indeed, gauge cancellations must be kept in this case between $q^2_i$ and the denominators of the loop functions. 
This is achieved by replacing $q^2_i$ with $\qbar^2_i$, defined in Eq.~(\ref{eq:split}), and performing the {\em same} subtraction in the integrals\footnote{Integrals involving  powers of $\mu^2$ are called {\em extra integrals}.} containing $\mu^2$ {\em as if} $\mu^2= q^\alpha_i q^\beta_i$.
For instance, one {\em defines}, in analogy with Eq.~(\ref{eq:tensor}),
\bqa
\label{eq:extraint}
\int [d^4q] \frac{\mu^2}{(\qbar^2-M^2)^3} &\equiv& \lim_{\mu \to 0}
\int_{\rm R} d^4q \left(
\frac{\mu^2}{(\qbar^2-M^2)^3} -\left[\frac{\mu^2}{\qbar^6}\right]
         \right) = \frac{i \pi^2}{2}\nl
&=& \lim_{\mu \to 0} \mu^2 \,\int d^4q \left(
   \frac{M^2}{\qbar^6(\qbar^2-M^2)}
 + \frac{M^2}{\qbar^4(\qbar^2-M^2)^2}
 + \frac{M^2}{\qbar^2(\qbar^2-M^2)^3}
         \right),\nl
\eqa
which preserves the simplification
\bqa
\label{eq:canc}
\int [d^4q] \frac{\qbar^2-M^2}{(\qbar^2-M^2)^3} &=& 
\lim_{\mu \to 0}
\int_{\rm R} d^4q \left(
\frac{\qbar^2-M^2}{(\qbar^2-M^2)^3} -\left[\frac{\qbar^2}{\qbar^6}\right]
         \right) \nl
&=&
\int [d^4q] \frac{1}{(\qbar^2-M^2)^2}.
\eqa

A second fundamental property of FDR integrals is shift invariance.
For instance,
\bqa
\label{eq:shift}
\int [d^4q] \frac{1}{(\qbar^2-M^2)^2}= 
\int [d^4q] \frac{1}{((q+p)^2-M^2-\mu^2)^2}
\eqa
is immediately manifest if DR is used as an ${\rm R}$ regulator in Eq.~(\ref{eq:1loopFDRint}).\footnote{See appendix A of~\cite{Donati:2013voa} for more details on shift invariance.}

FDR integration can be defined at any loop order. Partial fraction identities may  be used to split an $\ell$-loop integrand $J(q_1,\ldots,q_\ell)$ into its UV divergent part plus terms which are integrable in four dimensions
\bqa
\label{eq:splitting}
J(q_1,\ldots,q_\ell) = \left[J_{\rm INF}(q_1,\ldots,q_\ell)\right] + J_{\rm F}(q_1,\ldots,q_\ell),
\eqa
where $\left[J_{\rm INF}(q_1,\ldots,q_\ell)\right]$ collects all the subtracted integrands. In the multi-loop case, legal subtraction terms are also factorizable combinations of lower order divergent integrands times finite ones. For instance, a two-loop subtracted integrand can have the form
\bqa
\left[\frac{q_1^\alpha q_1^\beta}{\qbar_1^6}\right] \frac{1}{(\qbar^2_2-M^2)^3}.
\eqa
Thus, the FDR integral over $J(q_1,\ldots,q_\ell)$ reads
\bqa
\int [d^4 q_1] \cdots [d^4 q_\ell] J(q_1,\ldots,q_\ell)= 
\lim_{\mu \to 0} 
\int d^4 q_1 \cdots d^4 q_\ell
J_{\rm F}(q_1,\ldots,q_\ell),
\eqa
where, as discussed in the one-loop example, the replacement $\mu \to \mur$ is understood whenever powers of $\ln \mu^2$ appear. Finally, the generalizations of Eqs.~(\ref{eq:canc}) and~(\ref{eq:shift}) read
\bqa
\label{eq:canc1}
\int [d^4q_1] \ldots [d^4q_\ell]\, \frac{\qbar^2_i-m^2_i}{ (\qbar^2_i-m^2_i)^m \ldots} = 
\int [d^4q_1] \ldots [d^4q_\ell]\, \frac{1}{ (\qbar^2_i-m^2_i)^{m-1}\ldots}
\eqa
and
\bqa
\label{eq:shift1}
\int [d^4q_1] \ldots [d^4q_\ell]\,  J(q_1,\ldots, q_\ell) =\int [d^4q_1] \ldots [d^4q_\ell]\,  J(q_1+p_1,\ldots, q_\ell+p_\ell), 
\eqa
respectively.

The fact that the previous two Equations do not contain any reference to $\left[J_{\rm INF}\right]$ implies that subtracted integrands never play a role. Indeed Eqs.~(\ref{eq:canc1}) and~(\ref{eq:shift1}) state that manipulations in FDR integrands are allowed as if they were integrands of convergent integrals and that gauge cancellations automatically occur.
For instance, by using
$q^\alpha q^\beta \to \frac{g^{\alpha \beta}}{4}\,q^2$ and
$q^2= (\bar q^2-M^2)+M^2 + \mu^2$, the tensor reduction of 
Eq.~(\ref{eq:tensor}) reads
\bqa
\int [d^4q] \frac{q^\alpha q^\beta}{(\qbar^2-M^2)^3} &=& 
\frac{g^{\alpha \beta}}{4} \left(
    \int [d^4q] \frac{1}{(\qbar^2-M^2)^2}
+M^2\int [d^4q] \frac{1}{(\qbar^2-M^2)^3} \right. \nl
&&+ \left.    \int [d^4q] \frac{\mu^2}{(\qbar^2-M^2)^3}
\right),
\eqa
where the gauge symmetry preserving constant is generated by the last term. 
\section{Integration by parts}
\label{IBP}
In this Section, I recall the origin of the IBP identities~\cite{Tkachov:1981wb,Chetyrkin:1981qh} among integrals defined in DR. This serves as a basis for the extension discussed in Sec.~\ref{IBPinFDR}. 

The validity of the IBP relations in DR relies on the following two properties:
\bqa
\label{eq:2cond}
\begin{tabular}{l}
$\bullet$ Integration is the inverse process of differentiation; \\
$\bullet$ Surface terms never contribute if the integrals are evaluated in $n$ dimensions.
\end{tabular}
\eqa
Thus, one can write
\bqa
\label{eq:ibpdef}
0= \mur^{-\ell \epsilon} \int d^n q_1 \cdots d^n q_\ell \frac{\partial}{\partial q^\alpha_i}
\frac{v^\alpha}{D^{\nu_1}_1 \cdots D^{\nu_m}_m},
\eqa
where $D_1 \cdots D_m$ are loop propagators, $\nu_i$ are generic powers and $v^\alpha$ is a vector made of loop and/or external  momenta.
Acting with $\frac{\partial}{\partial q^\alpha_i}$ generates a sum of $s$ integrands
\bqa
\label{eq:jr}
\frac{\partial}{\partial q^\alpha_i}
\frac{v^\alpha}{D^{\nu_1}_1 \cdots D^{\nu_m}_m}
= \sum_{r=1}^s J_r(n,q_1,\ldots,q_\ell),
\eqa
where the $n$ among the arguments of  $J_r$ denotes a possible
dependence on the space-time dimensionality
\bqa
n= \frac{\partial q^\alpha_i}{\partial q^\alpha_i}.
\eqa
Integrating Eq.~(\ref{eq:jr}) gives a relation among the $s$ integrals
over the $J_r$
\bqa
\sum_{r=1}^s \mur^{-\ell \epsilon} \int d^n q_1 \cdots d^n q_\ell 
 J_r(n,q_1,\ldots,q_\ell) = 0.
\eqa

A one-loop example is given by
\bqa
0= \mur^{-\epsilon} \int d^n q   \frac{\partial}{\partial q^\alpha} \frac{q^\alpha}{D_0 D_1}, 
\eqa
with $D_0= q^2-m^2_0$ and $D_1= (q+p)^2-m^2_1$. Taking the derivative produces the identity
\bqa
\label{eq:eqxn1}
\mur^{-\epsilon}
\int d^n q 
\left\{
   \frac{n}{D_0 D_1}
-2 \frac{q^2}{D^2_0 D_1}
-2 \frac{q^2+(q\cdot p)}{D_0 D^2_1}
\right\} = 0.
\eqa
As a  two-loop case consider
\bqa
0= \mur^{-2 \epsilon} \int d^n q_1 d^n q_2   
\frac{\partial}{\partial q_1^\alpha} \frac{q_1^\alpha q_1^\beta q_1^\gamma}{D^3_1 D_2 D_{12}}, 
\eqa
with $D_i= q_i^2-m^2_i$ and $q_{12}= q_1+q_2$. Differentiating with respect to
$q_1^\alpha$ gives
\bqa
\label{eq:eqxn2}
\mur^{-2 \epsilon}
\int d^n q_1 d^n q_2 \,
q_1^\beta q_1^\gamma
\left\{
  \frac{2+n}{D^3_1 D_2 D_{12}}
-6\frac{q^2_1}{D^4_1 D_2 D_{12}}
-2 \frac{(q_1 \cdot q_{12})}{D^3_1 D_2 D^2_{12}}
\right\} = 0.
\eqa
\section{IBP identities in FDR}
\label{IBPinFDR}
FDR integration does not fulfill, in an explicit way, the two conditions in
Eq.~(\ref{eq:2cond}). In fact, it is four-dimensional and
 not defined as the inverse of differentiation. 
In spite of this, IBP identities hold in FDR. More explicitly, let
\bqa
\frac{\partial}{\partial q^\alpha_i} J^\alpha(q_1,\ldots,q_\ell)
\eqa
be the derivative with respect to the $i^{th}$ loop momentum of the integrand of an FDR $\ell$-loop function.
Acting with $\frac{\partial}{\partial q^\alpha_i}$ on $J^\alpha$  gives, in analogy with Eq.~(\ref{eq:jr}), a sum of $s$ integrands $J_r$
\bqa
\label{eq:int3}
\frac{\partial}{\partial q^\alpha_i} J^\alpha(q_1,\ldots,q_\ell) = \sum_{r=1}^s J_r(4,q_1,\ldots,q_\ell),
\eqa
where
\bqa
4= \frac{\partial q^\alpha_i}{\partial q^\alpha_i}.
\eqa
In this Section I prove that
\bqa
\label{eq:IBP}
0= \int [d^4 q_1] \cdots [d^4 q_\ell] \frac{\partial}{\partial q^\alpha_i}
J^\alpha(q_1,\ldots,q_l) 
=
\sum_{r=1}^{s}
\int [d^4 q_1] \cdots [d^4 q_\ell] 
 J_r(4,q_1,\ldots,q_\ell),
\eqa
which demonstrates that four-dimensional IBP identities can be used in FDR integrals.

The proof of Eq.~(\ref{eq:IBP}) is as follows.
According to Eq.~(\ref{eq:splitting})
\bqa
\int [d^4 q_1] \cdots [d^4 q_\ell] \frac{\partial}{\partial q^\alpha_i}
J^\alpha(q_1,\ldots,q_l)
\eqa  
is defined by splitting
\bqa
\label{eq:splitting1}
\frac{\partial}{\partial q^\alpha_i}J^\alpha(q_1,\ldots,q_\ell) = 
\frac{\partial}{\partial q^\alpha_i}\Bigl(\left[J^\alpha_{\rm INF}(q_1,
\ldots,q_\ell)\right] + J^\alpha_{\rm F}(q_1,\ldots,q_\ell)\Bigr)
\eqa
in such a way that
\bqa
\frac{\partial}{\partial q^\alpha_i} J^\alpha_{\rm F}(q_1,\ldots,q_\ell)
\eqa
is integrable in four dimensions.
Then
\bqa
\label{eq:int1}
\int [d^4 q_1] \cdots [d^4 q_\ell] \frac{\partial}{\partial q^\alpha_i}
J^\alpha(q_1,\ldots,q_l) \equiv \lim_{\mu \to 0} 
\int d^4 q_1 \cdots d^4 q_\ell 
 \frac{\partial}{\partial q^\alpha_i}
J^\alpha_{\rm F}(q_1,\ldots,q_l) = 0,
\eqa
where the r.h.s. vanishes because it is the integral of a total derivative.
In addition, Eq.~(\ref{eq:splitting1}) allows one to rewrite
\bqa
\label{eq:int2}
&&
\lim_{\mu \to 0} 
\int d^4 q_1 \cdots d^4 q_\ell 
 \frac{\partial}{\partial q^\alpha_i}
J^\alpha_{\rm F}(q_1,\ldots,q_l) \nl
&&~~~= \lim_{\mu \to 0} \mur^{-\ell \epsilon} 
\int d^n q_1 \cdots d^n q_\ell 
\left(
\frac{\partial}{\partial q^\alpha_i} J^\alpha(q_1,\ldots,q_\ell)-
\frac{\partial}{\partial q^\alpha_i} \left[J^\alpha_{\rm INF}(q_1,\ldots,q_\ell)\right]
\right),
\eqa
where the two separately UV divergent integrals in the r.h.s. are computed in DR. The action of the derivative on $J^\alpha$  gives  
\bqa
\label{eq:int33}
\int d^n q_1 \cdots d^n q_\ell 
\frac{\partial}{\partial q^\alpha_i} J^\alpha(q_1,\ldots,q_\ell) &=& 
\int d^n q_1 \cdots d^n q_\ell \sum_{r=1}^s J_r(n,q_1,\ldots,q_\ell),
\eqa
which differs from Eq.~(\ref{eq:int3}) in that
\bqa
n= \frac{\partial q^\alpha_i}{\partial q^\alpha_i}.
\eqa
On the other hand, acting with $\frac{\partial}{\partial q^\alpha_i}$ on
$\left[J^\alpha_{\rm INF}\right]$ generates, in general, $s^\prime$ terms
\bqa
\label{eq:int44}
\int d^n q_1 \cdots d^n q_\ell 
\frac{\partial}{\partial q^\alpha_i} \left[J^\alpha_{\rm INF}(q_1,\ldots,q_\ell)\right] &=& 
\int d^n q_1 \cdots d^n q_\ell \sum_{r=1}^{s^\prime} K_r(n,q_1,\ldots,q_\ell).
\eqa
However, owing to Eq.~(\ref{eq:splitting1}), the r.h.s. of Eq.~(\ref{eq:int44}) has to match, term by term, the INF part of  Eq.~(\ref{eq:int33}), so that it is possible to recast it as follows
\bqa
\int d^n q_1 \cdots d^n q_\ell \sum_{r=1}^{s^\prime} K_r(n,q_1,\ldots,q_\ell) = 
\int d^n q_1 \cdots d^n q_\ell \sum_{r=1}^{s}      \left[J_{r,{\rm INF}}(n,q_1,\ldots,q_\ell)\right].
\eqa
Therefore
\bqa
\label{eq:int4}
&&\lim_{\mu \to 0} 
\int d^4 q_1 \cdots d^4 q_\ell 
 \frac{\partial}{\partial q^\alpha_i}
J^\alpha_{\rm F}(q_1,\ldots,q_l) \nl
&&~~~= \lim_{\mu \to 0} \mur^{-\ell \epsilon}  
\int d^n q_1 \cdots d^n q_\ell 
\sum_{r=1}^{s}  \Bigl(
    J_r(n,q_1,\ldots,q_\ell)-\left[J_{r,{\rm INF}}(n,q_1,\ldots,q_\ell)\right]
              \Bigr).
\eqa
The r.h.s. of Eq.~(\ref{eq:int4}) is nothing but the sum of the $s$ FDR integrals over the  $J_r$ integrands, and --\,since the difference is UV convergent\,-- one can set $n=4$. Thus Eq.~(\ref{eq:IBP}) follows from Eqs.~(\ref{eq:int1}) and (\ref{eq:int4}).

Notice that, as in the case of Eqs.~(\ref{eq:canc1}) and~(\ref{eq:shift1}), no reference is made, in Eq.~(\ref{eq:IBP}), to the explicit form of $\left[J^\alpha_{\rm INF}\right]$ or $J^\alpha_{\rm F}$. One directly differentiates $J^\alpha$ with respect to $q_i^\alpha$, as in Eq.~(\ref{eq:int3}). 
For instance, the FDR counterparts of Eqs.~(\ref{eq:eqxn1}) and~(\ref{eq:eqxn2})  read
\bqa
\label{eq:eqx41}
\int [d^4 q] \frac{\partial}{\partial q^\alpha} \frac{q^\alpha}{\bar D_0 \bar D_1} = \int [d^4 q] 
\left\{
   \frac{4}{\bar D_0 \bar D_1}
-2 \frac{q^2}{\bar D^2_0 \bar D_1}
-2 \frac{q^2+(q\cdot p)}{\bar D_0 \bar D^2_1}
\right\} = 0
\eqa
and
\bqa
\label{eq:eqx42}
&&
\int [d^4 q_1] [d^4 q_2]   
\frac{\partial}{\partial q_1^\alpha} \frac{q_1^\alpha q_1^\beta q_1^\gamma}{\bar D^3_1\bar  D_2 \bar D_{12}} 
\nl 
&&~~~
= \int [d^4 q_1] [d^4 q_2] \,
q_1^\beta q_1^\gamma
\left\{
 \frac{6}{\bar D^3_1 \bar D_2 \bar D_{12}}
-\frac{6 q^2_1}{\bar D^4_1 \bar D_2 \bar D_{12}}
-2 \frac{(q_1 \cdot q_{12})}{\bar D^3_1 \bar D_2 \bar D^2_{12}}
\right\} = 0,
\eqa
respectively, where $\bar D_i= D_i-\mu^2$.

A crucial difference from DR is that loop momenta squared 
{\em do not directly cancel denominators}. For example, extra integrals [such as the one given in Eq.~(\ref{eq:extraint})] are created when the identity $q^2= \bar D_0+m^2_0+\mu^2$ is used to scalarize Eq.~(\ref{eq:eqx41}).
These extra terms play the role of the $\epsilon/\epsilon$ constant generated by the presence of $n= 4 + \epsilon$ in Eq.~(\ref{eq:eqxn1}). 
The advantage of Eq.~(\ref{eq:eqx41}) versus Eq.~(\ref{eq:eqxn1}) is that 
{\em all} integrals contributing to the IBP identity appear on the same footing, without expanding in $\epsilon$. This is why simpler IBP based algorithms are expected in FDR.

Although the proof of Eq.~(\ref{eq:IBP}) given here is completely general, it is instructive to elucidate it further with a couple of examples. With this aim,
the explicit derivation of Eqs.~(\ref{eq:eqx41}) and~(\ref{eq:eqx42}) is presented in Appendix~\ref{appb}. 
\section{Conclusions}
\label{concl}
Eq.~(\ref{eq:IBP}) is the main result of this paper. It states that four-dimensional IBP identities can be exploited to establish relations among multi-loop FDR integrals. Owing to the fact that IBP based techniques are extensively employed to identify and determine MIs in higher-order QFT calculations, this outcome paves the way for the use of FDR in multi-loop computations.

Since FDR loop calculus is carried out in four dimensions and without an explicit use of UV counterterms, a reduction in complexity is envisaged compared with IBP algorithms based on DR. 
This last aspect will be investigated more in detail in future publications.

\acknowledgments

This research was supported by the European Commission through contracts ERC-2011-AdG No 291377 (LHCtheory) and PITN-GA-2012-316704 (HIGGSTOOLS).
I also thank the projects FPA2011-22398 (LHC@NLO), FPA2013-47836-C3-1-P and P10-FQM-6552.
Work supported by the Munich Institute for Astro- and Particle Physics (MIAPP) of the DFG cluster of excellence ``Origin and Structure of the Universe''.

\appendix
\section{Two examples}
\label{appb}
In this appendix, I derive Eqs.~(\ref{eq:eqx41}) and~(\ref{eq:eqx42})
along the lines of the general proof given in Sec~\ref{IBPinFDR}.

Partial fractioning generates the identity
\bqa
\frac{q^\alpha}{\bar D_0 \bar D_1} = 
\left[\frac{q^\alpha}{\qbar^4} \right]
+ q^\alpha 
\left( 
\frac{M^2-p^2-2(q \cdot p)}{\qbar^4 \bar D_1}
+ \frac{M^2}{\qbar^2 \bar D_0 \bar D_1}
\right).
\eqa
Differentiating the second term with respect to $q^\alpha$ produces
a function of $q$ integrable in four dimensions.
Thus
\bqa
0 &=& \lim_{\mu \to 0} \int d^4q \frac{\partial}{\partial q^\alpha}
\left\{
 q^\alpha 
\left( 
\frac{M^2-p^2-2(q \cdot p)}{\qbar^4 \bar D_1}
+ \frac{M^2}{\qbar^2 \bar D_0 \bar D_1}
\right)
\right\} \nl
 &=&
\lim_{\mu \to 0}
\mur^{-\epsilon}
\int d^n q
\frac{\partial}{\partial q^\alpha} 
\left\{
\frac{q^\alpha}{\bar D_0 \bar D_1}
-
\left[\frac{q^\alpha}{\qbar^4} \right]
\right\}.
\eqa
But
$
\frac{\partial}{\partial q^\alpha} 
\frac{q^\alpha}{\bar D_0 \bar D_1}
$ in $n$ dimensions
is given by the integrand in Eq.~(\ref{eq:eqxn1}), with 
$D_i \to \bar D_i$, and
\bqa
\frac{\partial}{\partial q^\alpha} 
\frac{q^\alpha}{\qbar^4}
=
\frac{n}{\qbar^4}
-4 \frac{q^2}{\qbar^6},
\eqa
from which Eq.~(\ref{eq:eqx41}) follows
\bqa
0 &=&  \lim_{\mu \to 0} \mur^{-\epsilon} 
\int d^n q
\left\{
n 
\left(
      \frac{1}{\bar D_0 \bar D_1}
-\left[\frac{1}{\bar \qbar^4}\right]
\right)
- 2 q^2
\left(
       \frac{1}{\bar D^2_0 \bar D_1}
-\left[\frac{1}{\bar \qbar^6}\right]
\right) \right. \nl
 && \left.- 2 q^2
\left(
       \frac{1}{\bar D_0 \bar D^2_1}
-\left[\frac{1}{\bar \qbar^6}\right]
\right)
-2 \frac{(q \cdot p)}{\bar D_0 \bar D^2_1}
\right\} \nl
&=& \int [d^4 q] 
\left\{
   \frac{4}{\bar D_0 \bar D_1}
-2 \frac{q^2}{\bar D^2_0 \bar D_1}
-2 \frac{q^2+(q\cdot p)}{\bar D_0 \bar D^2_1}
\right\}.
\eqa

As for Eq.~(\ref{eq:eqx42}), the needed identity is
\bqa
\label{eq:split2}
\frac{q^\alpha_1 q^\beta_1 q^\gamma_1}{\dbar^3_1 \dbar_2 \dbar_{12}} &=&
q^\alpha_1 q^\beta_1 q^\gamma_1
\left\{
\left[\frac{1}{\qbar^6_1 \qbar^2_2 \qbar^2_{12}}  \right]
   +\left(
            \frac{m^2_1}{\bar D^3_1 \bar{q}_1^2}
           +\frac{m^2_1}{\bar D^2_1 \bar{q}_1^4}
           +\frac{m^2_1}{\bar D_1   \bar{q}_1^6}
    \right)
\left[\frac{1}{\qbar^4_2}  \right]
\right\} + J_{\rm F}^{\alpha \beta \gamma},
\eqa
where
\bqa
J_{\rm F}^{\alpha \beta \gamma} &=&
q^\alpha_1 q^\beta_1 q^\gamma_1
\left\{
\frac{1}{\dbar^3_1 \bar{q}_2^2 \dbar_{12}} 
\left(
\frac{m^2_2}{\dbar_2}
+
\frac{m^2_{12}}{\bar{q}_{12}^2}
\right)
          -\left(
            \frac{m^2_1}{\bar D^3_1 \bar{q}_1^2}
           +\frac{m^2_1}{\bar D^2_1 \bar{q}_1^4}
           +\frac{m^2_1}{\bar D_1   \bar{q}_1^6}
           \right)
\frac{q_1^2+2 (q_1 \cdot q_2)}{\bar{q}_2^4 \bar{q}_{12}^2}
\right\} \nl
\eqa
is such that the four dimensional two-loop integral over $\frac{\partial}{\partial q_1^\alpha} J_{\rm F}^{\alpha \beta \gamma}$ is convergent.
The derivative with respect to $q_1^\alpha$ of 
the l.h.s. of Eq.~(\ref{eq:split2}) can be read from the integrand in Eq.~(\ref{eq:eqxn2}).  Furthermore
\bqa
\frac{\partial}{\partial q_1^\alpha} 
\frac{q_1^\alpha q_1^\beta q_1^\gamma}{\qbar^6_1 \qbar^2_2 
\qbar^2_{12}}= 
q_1^\beta q_1^\gamma
\left\{
    \frac{2+n}{\qbar^6_1 \qbar^2_2 \qbar^2_{12}}
-6\frac{q^2_1}{\qbar^8_1 \qbar^2_2 \qbar^2_{12}}
-2 \frac{(q_1 \cdot q_{12})}{\qbar^6_1 \qbar^2_2 \qbar^4_{12}}
\right\},
\eqa 
and
\bqa
\frac{\partial}{\partial q_1^\alpha} 
q_1^\alpha q_1^\beta q_1^\gamma
    \left(
            \frac{m^2_1}{\bar D^3_1 \bar{q}_1^2}
           +\frac{m^2_1}{\bar D^2_1 \bar{q}_1^4}
           +\frac{m^2_1}{\bar D_1   \bar{q}_1^6}
    \right)
&=&
q_1^\beta q_1^\gamma
\left\{
(2+n)
    \left(
            \frac{m^2_1}{\bar D^3_1 \bar{q}_1^2}
           +\frac{m^2_1}{\bar D^2_1 \bar{q}_1^4}
           +\frac{m^2_1}{\bar D_1   \bar{q}_1^6}
    \right) \right. \nl
&&\left. -6 q^2_1
    \left(
            \frac{m^2_1}{\bar D^4_1 \bar{q}_1^2}
           +\frac{m^2_1}{\bar D^3_1 \bar{q}_1^4}
           +\frac{m^2_1}{\bar D^2_1 \bar{q}_1^6}
           +\frac{m^2_1}{\bar D_1   \bar{q}_1^8}
    \right)
\right\}. \nl
\eqa
Therefore, 
\bqa
0 &=&  \lim_{\mu \to 0} \int d^4q_1  d^4q_2 
\frac{\partial}{\partial q_1^\alpha} 
J_{\rm F}^{\alpha \beta \gamma} \nl
&=&  
(2+n) \lim_{\mu \to 0} \mur^{-2 \epsilon} \int d^nq_1 \int d^nq_2 
\,q_1^\beta q_1^\gamma\,
\left\{ 
  \frac{1}{\dbar^3_1 \dbar_2 \dbar_{12}}
-\left[\frac{1}{\qbar^6_1 \qbar^2_2 \qbar^2_{12}}  \right]
\right. \nl
&&
- \left. \left(
            \frac{m^2_1}{\bar D^3_1 \bar{q}_1^2}
           +\frac{m^2_1}{\bar D^2_1 \bar{q}_1^4}
           +\frac{m^2_1}{\bar D_1   \bar{q}_1^6}
  \right)
\left[\frac{1}{\qbar^4_2}  \right]
\right\}\nl
&& -6 
\lim_{\mu \to 0} \mur^{-2 \epsilon} \int d^nq_1 \int d^nq_2 
\,q_1^\beta q_1^\gamma\,q^2_1
\left\{ 
  \frac{1}{\dbar^4_1 \dbar_2 \dbar_{12}}
-\left[\frac{1}{\qbar^8_1 \qbar^2_2 \qbar^2_{12}}  \right]
\right.  \nl
&& 
- \left.\left(
            \frac{m^2_1}{\bar D^4_1 \bar{q}_1^2}
           +\frac{m^2_1}{\bar D^3_1 \bar{q}_1^4}
           +\frac{m^2_1}{\bar D^2_1 \bar{q}_1^6}
           +\frac{m^2_1}{\bar D_1   \bar{q}_1^8}
  \right)
\left[\frac{1}{\qbar^4_2}  \right]
\right\} \nl
&&-2 
\lim_{\mu \to 0} \mur^{-2 \epsilon} \int d^nq_1 \int d^nq_2 
\,q_1^\beta q_1^\gamma\,
(q_1 \cdot q_{12})
\left\{ 
  \frac{1}{\dbar^3_1 \dbar_2 \dbar^2_{12}}
-\left[\frac{1}{\qbar^6_1 \qbar^2_2 \qbar^4_{12}}  \right]
\right\} \nl
&=& \int [d^4 q_1] [d^4 q_2] \,
q_1^\beta q_1^\gamma
\left\{
 \frac{6}{\bar D^3_1 \bar D_2 \bar D_{12}}
-\frac{6 q^2_1}{\bar D^4_1 \bar D_2 \bar D_{12}}
-2 \frac{(q_1 \cdot q_{12})}{\bar D^3_1 \bar D_2 \bar D^2_{12}}
\right\}.
\eqa
\bibliography{paper}{}

\providecommand{\href}[2]{#2}\begingroup\raggedright\begin{thebibliography}{10}

\bibitem{Collins:1984xc}
J.~C. Collins, {\em Renormalization}.
\newblock Cambridge University Press, 1984.

\bibitem{'tHooft:1972fi}
G.~'t~Hooft and M.~Veltman, {\it {Regularization and Renormalization of Gauge
  Fields}},  {\em Nucl.Phys.} {\bf B44} (1972) 189--213.

\bibitem{Bogoliubov:1957gp}
N.~Bogoliubov and O.~a. Parasiuk, {\it {On the Multiplication of the causal
  function in the quantum theory of fields}},  {\em Acta Math.} {\bf 97} (1957)
  227--266.

\bibitem{Hepp:1966eg}
K.~Hepp, {\it {Proof of the Bogolyubov-Parasiuk theorem on renormalization}},
  {\em Commun.Math.Phys.} {\bf 2} (1966) 301--326.

\bibitem{Zimmermann:1969jj}
W.~Zimmermann, {\it {Convergence of Bogolyubov's method of renormalization in
  momentum space}},  {\em Commun.Math.Phys.} {\bf 15} (1969) 208--234.

\bibitem{Pittau:2012zd}
R.~Pittau, {\it {A four-dimensional approach to quantum field theories}},  {\em
  JHEP} {\bf 1211} (2012) 151, [\href{http://xxx.lanl.gov/abs/1208.5457}{{\tt
  arXiv:1208.5457}}].

\bibitem{Pittau:2013qla}
R.~Pittau, {\it {QCD corrections to $H \to gg$ in FDR}},
  \href{http://xxx.lanl.gov/abs/1307.0705}{{\tt arXiv:1307.0705}}.

\bibitem{Donati:2013iya}
A.~Donati and R.~Pittau, {\it {Gauge invariance at work in FDR: H $\to$ gamma
  gamma}},  \href{http://xxx.lanl.gov/abs/1302.5668}{{\tt arXiv:1302.5668}}.

\bibitem{Donati:2013voa}
A.~M. Donati and R.~Pittau, {\it {FDR, an easier way to NNLO calculations: a
  two-loop case study}},  {\em Eur.Phys.J.} {\bf C74} (2014) 2864,
  [\href{http://xxx.lanl.gov/abs/1311.3551}{{\tt arXiv:1311.3551}}].

\bibitem{Pittau:2013ica}
R.~Pittau, {\it {On the predictivity of the non-renormalizable quantum field
  theories}},  \href{http://xxx.lanl.gov/abs/1305.0419}{{\tt arXiv:1305.0419}}.

\bibitem{Tkachov:1981wb}
F.~Tkachov, {\it {A Theorem on Analytical Calculability of Four Loop
  Renormalization Group Functions}},  {\em Phys.Lett.} {\bf B100} (1981)
  65--68.

\bibitem{Chetyrkin:1981qh}
K.~Chetyrkin and F.~Tkachov, {\it {Integration by Parts: The Algorithm to
  Calculate beta Functions in 4 Loops}},  {\em Nucl.Phys.} {\bf B192} (1981)
  159--204.

\bibitem{Laporta:2001dd}
S.~Laporta, {\it {High precision calculation of multiloop Feynman integrals by
  difference equations}},  {\em Int.J.Mod.Phys.} {\bf A15} (2000) 5087--5159,
  [\href{http://xxx.lanl.gov/abs/hep-ph/0102033}{{\tt hep-ph/0102033}}].

\bibitem{Marquard:2008qn}
P.~Marquard, {\it {New Methods for the Calculation of Multi-Loop Amplitudes}},
  \href{http://xxx.lanl.gov/abs/0810.3620}{{\tt arXiv:0810.3620}}.

\bibitem{Kotikov:1990kg}
A.~Kotikov, {\it {Differential equations method: New technique for massive
  Feynman diagrams calculation}},  {\em Phys.Lett.} {\bf B254} (1991) 158--164.

\bibitem{Remiddi:1997ny}
E.~Remiddi, {\it {Differential equations for Feynman graph amplitudes}},  {\em
  Nuovo Cim.} {\bf A110} (1997) 1435--1452,
  [\href{http://xxx.lanl.gov/abs/hep-th/9711188}{{\tt hep-th/9711188}}].

\bibitem{Gehrmann:1999as}
T.~Gehrmann and E.~Remiddi, {\it {Differential equations for two loop four
  point functions}},  {\em Nucl.Phys.} {\bf B580} (2000) 485--518,
  [\href{http://xxx.lanl.gov/abs/hep-ph/9912329}{{\tt hep-ph/9912329}}].

\bibitem{Argeri:2007up}
M.~Argeri and P.~Mastrolia, {\it {Feynman Diagrams and Differential
  Equations}},  {\em Int.J.Mod.Phys.} {\bf A22} (2007) 4375--4436,
  [\href{http://xxx.lanl.gov/abs/0707.4037}{{\tt arXiv:0707.4037}}].

\bibitem{Smirnov:2012gma}
V.~A. Smirnov, {\it {Analytic tools for Feynman integrals}},  {\em Springer
  Tracts Mod.Phys.} {\bf 250} (2012) 1--296.

\bibitem{Papadopoulos:2014lla}
C.~G. Papadopoulos, {\it {Simplified differential equations approach for Master
  Integrals}},  {\em JHEP} {\bf 1407} (2014) 088,
  [\href{http://xxx.lanl.gov/abs/1401.6057}{{\tt arXiv:1401.6057}}].

\bibitem{Henn:2013pwa}
J.~M. Henn, {\it {Multiloop integrals in dimensional regularization made
  simple}},  {\em Phys.Rev.Lett.} {\bf 110} (2013), no.~25 251601,
  [\href{http://xxx.lanl.gov/abs/1304.1806}{{\tt arXiv:1304.1806}}].

\bibitem{Argeri:2014qva}
M.~Argeri, S.~Di~Vita, P.~Mastrolia, E.~Mirabella, J.~Schlenk, et~al., {\it
  {Magnus and Dyson Series for Master Integrals}},  {\em JHEP} {\bf 1403}
  (2014) 082, [\href{http://xxx.lanl.gov/abs/1401.2979}{{\tt
  arXiv:1401.2979}}].

\bibitem{DiVita:2014pza}
S.~Di~Vita, P.~Mastrolia, U.~Schubert, and V.~Yundin, {\it {Three-loop master
  integrals for ladder-box diagrams with one massive leg}},
  \href{http://xxx.lanl.gov/abs/1408.3107}{{\tt arXiv:1408.3107}}.

\bibitem{Passarino:1978jh}
G.~Passarino and M.~Veltman, {\it {One Loop Corrections for e+ e- Annihilation
  Into mu+ mu- in the Weinberg Model}},  {\em Nucl.Phys.} {\bf B160} (1979)
  151.

\bibitem{Ossola:2006us}
G.~Ossola, C.~G. Papadopoulos, and R.~Pittau, {\it {Reducing full one-loop
  amplitudes to scalar integrals at the integrand level}},  {\em Nucl.Phys.}
  {\bf B763} (2007) 147--169,
  [\href{http://xxx.lanl.gov/abs/hep-ph/0609007}{{\tt hep-ph/0609007}}].

\end{thebibliography}\endgroup
\bibliographystyle{JHEP}
\end{document}